\documentclass{PoS_arXiv}

\title{Charged Higgs Bosons in the LHCHXSWG}

\usepackage{amsmath,amssymb,amsfonts,color,graphicx,cite,color,soul}

\ShortTitle{LHCHXSWG: Charged Higgs}

\author{\speaker{S. Heinemeyer}\\ 
        Instituto de F\'isica de Cantabria (CSIC-UC), Santander, Spain\\
        E-mail: \email{Sven.Heinemeyer@cern.ch}}

\abstract{Searches for charged Higgs bosons are an integral part of current
  and future investigations at the LHC. The LHC Higgs Cross Section
  Working Group 
  (LHCHXSWG) was created to provide cross sections, branching ratios,
  analysis strategies etc.\ for Higgs boson searches at the LHC.
  We briefly review progress and results for charged Higgs bosons in and
  for the LHCHXSWG.}

\FullConference{Prospects for Charged Higgs Discovery at Colliders - CHARGED
  2014,\\ 
		16-18 September 2014\\
		Uppsala University, Sweden}

\include{paperdef}
\graphicspath{{figs/}}

\newcommand{\simMH}{125}

\begin{document}


\section{Introduction}

A major goal of the 
particle physics program at the high energy frontier,
currently being pursued at the CERN Large Hadron Collider (LHC),
is to unravel the nature of electroweak symmetry breaking (EWSB).
While the existence of the massive electroweak gauge bosons ($W^\pm,Z$),
together with the successful description of their behavior by
non-abelian gauge theory, 
requires some form of EWSB to be present in nature, 
the underlying dynamics remained unknown for several decades. 
An appealing theoretical suggestion for such dynamics is the Higgs mechanism
\cite{higgs-mechanism}, which 
implies the existence of one or more 
Higgs bosons (depending on the specific model considered).
Therefore, the search for a Higgs boson was considered a major cornerstone
in the physics program of the LHC.

The spectacular discovery of a Higgs-like particle 
with a mass around $\MH \simeq \simMH \gev$, which has been announced
by ATLAS \cite{ATLASdiscovery} and CMS~\cite{CMSdiscovery}, marks a
milestone of an effort that has been ongoing for almost half a century
and opens up a new era of particle physics.  
Both ATLAS and CMS reported a clear excess in the two photon channel, as
well as in the $ZZ^{(*)}$ channel. The discovery was further 
corroborated, though not with high significance, by the
$WW^{(*)}$ channel and by the final Tevatron results~\cite{TevHiggsfinal}.
Latest ATLAS/CMS results, also for evidence on the Higgs decay into
fermions can be found in \citeres{ATLAS-Higgs-WWW,CMS-Higgs-WWW}. 

\medskip
Many theoretical models employing the Higgs mechanism in
order to account for electroweak symmetry breaking
have been studied in the literature, of which 
the most popular ones are the Standard Model (SM)~\cite{sm}  
and the Minimal Supersymmetric Standard Model (MSSM)~\cite{mssm}, 
The newly discovered particle can be interpreted as the SM Higgs boson.
The MSSM has a richer Higgs sector, containing two neutral $\cp$-even,
one neutral $\cp$-odd and two charged Higgs bosons. 
The newly discovered particle can also be interpreted as the light (or the
the heavy) $\cp$-even state~\cite{Mh125}. 
Among alternative theoretical models beyond the SM and the MSSM,
the most prominent are  
the (more general) Two Higgs Doublet Model (2HDM)~\cite{thdm,thdm-types}, 
non-minimal supersymmetric extensions of the SM 
(e.g.\ extensions of the MSSM by an extra singlet
superfield \cite{NMSSM-etc}), or models involving Higgs
triplets~\cite{triplet}. 
Many of these models not only predict more than one Higgs boson, but they
predict electrically charged Higgs bosons.

\medskip
The ATLAS and CMS analyses leading to the conclusion that (within the
uncertainties) the newly discovered particle can be interpreted as the SM
Higgs boson requires, besides the obvious experimental data, also precise
theory predictions for the SM Higgs boson cross section, branching ratios,
angular distributions as well as strategies how to extract certain
``measurements'' (e.g.\ coupling strength factors) from the data. 
In this respect it is crucial that ATLAS and CMS not only use predictions with
highest precision, but in particular that they use the {\em same} theory
predictions, the {\em same} strategies for the extraction of
``measurements''. Only then it is possible to readily compare ATLAS and
CMS results, and in the future combine them. 
To ensure this, in the year 2010 the 
``LHC Higgs Cross Section Group'' (LHCHXSWG)~\cite{LHCHXSWG-www1} was founded. 
This group, formed of theoretical and experimental physicists,
officially takes care of providing cross section and 
branching ratio predictions (including uncertainty evaluations), as well
as the strategies for the extraction of, e.g., coupling strength factors
from experimental data~\cite{YR1,YR2,YR3,HiggsRecommendation}. 
While initially the SM Higgs boson was in the
focus of the LHCHXSWG, soon also models beyond the SM (BSM) were
investigated (see also \citere{LHCHXSWG-www2}). In particular within the
MSSM cross sections and branching 
ratios for the extended Higgs sector have been evaluated, see, e.g.,
\citere{MSSMHiggsXS} for an example on the neutral Higgs production
cross sections. Latest results can be found at \citere{LHCHXSWG-www3}.

As discussed above, electrically charged Higgs bosons form a natural
part of many BSM models. 
The charged Higgs bosons of the MSSM (or a more general 2HDM)
have been searched at LEP, the Tevatron and the LHC, and
will be searched for (or hopefully analyzed at) a Linear Collider such
as ILC or CLIC. The LEP searches~\cite{ADLOchargedHiggs}
yielded a robust bound of
$\MHp \gsim 80 \gev$~\cite{LEPchargedHiggs}.
The Tevatron bounds~\cite{Tevcharged} are by now superseeded by the LHC
charged Higgs searches~\cite{LHCcharged}.
At the ILC, if the charged Higgs is in the
kinematical reach, a high-precision determination of the
charged Higgs boson properties will be
possible~\cite{ILC-TDR,MHpLHCILCnewer}.
Here, besides some basics, we briefly review activities and results
obtained within and for the LHCHXSWG regarding charged Higgs bosons, 
which will mainly concern the 2HDMs and the MSSM.



\bigskip
Within the 2HDM and the MSSM 
the main production channels of charged Higgs bosons at the LHC are
\begin{align}
\label{pp2Hpm}
pp \to t\bar t \; + \; X, \quad
t \bar t \to t \; H^- \bar b \mbox{~~or~~} H^+ b \; \bar t , \\
\label{gb2Hpm}
gb \to H^- t \mbox{~~or~~} g \bar b \to H^+ \bar t~ \quad \mbox{(5FS)}\,, \\
\label{gg2Hpm}
gg/q\bar q \to H^- t \bar b \mbox{~~or~~} 
gg/q\bar q \to H^+ \bar t b~ \quad \mbox{(4FS)}\,. 
\end{align}
The decay used in the analysis to detect the charged Higgs boson is
\begin{align}
H^\pm \; \to \; \tau \nu_\tau \; \to \; {\rm hadrons~}\nu_\tau. 
\label{Hdec}
\end{align}

The \ul{``light charged Higgs boson''} is characterized by $\MHp < \mt$. 
The main production channel is given in \refeq{pp2Hpm}. Close to
threshold also \refeq{gb2Hpm} contributes. The relevant (i.e.\
detectable) decay channel is given by \refeq{Hdec}.

The \ul{``heavy charged Higgs boson''} is characterized by $\MHp \gsim \mt$.
Here \refeq{gb2Hpm} in the ``five flavor scheme'' (5FS) and/or 
\refeq{gg2Hpm} in the ``four flavor scheme'' (4FS) gives the largest
contribution to the production cross section, and very close to 
threshold \refeq{pp2Hpm} can contribute somewhat. The relevant decay
channel is again given in \refeq{Hdec}.


\section{Charged Higgs bosons in 2HDMs}
\label{sec:2hdm}

The 2HDM can be classified in types I-IV~\cite{thdm-types}, where the
MSSM, see \refse{sec:mssm} at the tree-level contains a 2HDM type~II.
The relevant free (input) parameters are $\MHp$ and the ratio of the two
vacuum expectation values, $\tb \equiv v_2/v_1$.
Analyses at ATLAS and CMS
in the case of light charged Higgs bosons in the context of 2HDMs
evaluate the production cross section from $\si(pp \to t \bar t + X)$ as
evaluated in the SM at the NNLO level~\cite{ppttNNLO}. Limits are then
presented for $\br(t \to H^\pm b)$ as a function of the charged Higgs boson
mass,~$\MHp$. 

\medskip
For heavy charged Higgs bosons, $\MHpm \gsim \mt$,
associated production $pp \to tb H^\pm{\rm + X}$ is the
dominant production mode. 
Two different formalisms can be employed to calculate the cross
section for associated $tb H^\pm$ production.  In the
four-flavor scheme (4FS) with no $b$~quarks in the initial state,
the lowest-order QCD production processes are given in \refeq{gg2Hpm}.

On the other hand, potentially
large logarithms $\propto \ln(\mu_{\rm F}/\mb)$ (where $\mu_{\rm F}$
denotes the factorization scale), which arise from
the splitting of incoming gluons into nearly collinear $b \bar b$
pairs, can be summed to all orders in perturbation theory by
introducing bottom parton densities, i.e.\ in the 
five flavor scheme (5FS)~\cite{Barnett:1987jw}, see \refeq{gb2Hpm}.

To all orders in perturbation theory the four-
and five-flavor schemes are identical, but the way of ordering the
perturbative expansion is different, and the results do not match
exactly at finite order. For more details see \citere{YR2} and
references therein.
A simple and pragmatic formula for the combination of the 
four- and five-flavor scheme calculations of 
bottom-quark associated Higgs-boson production
has been suggested in \citere{Harlander:2011aa}, the so-called 
``Santander matching''. 
The main idea behind this matching scheme is the following:
The 4FS and 5FS calculations provide the unique description of the
cross section in the asymptotic limits $\MH/\mb \to 1$ and
$\MH/\mb \to \infty$, respectively (where $\MH$ denotes a
generic Higgs boson mass, i.e.\ the arguments are valid for the neutral
as well as for the charged Higgs production). 
The two approaches are combined in
such a way that they are given a weight, depending on the value
of the Higgs-boson mass. Since the difference between the 4FS and the 5FS
is logarithmic, the dependence of their relative importance
on $\MH$ should be controlled by a logarithmic term.
Consequently, the proposal for the ``Santander matching''
reads~\cite{Harlander:2011aa}, 
\begin{align}
\si^\text{matched} &= \frac{\si^\text{4FS} + t\,\si^\text{5FS}}{1+t}\,,
\quad \mbox{~with the weight $w$ defined as~} \quad
t = \ln\frac{\MH}{\mb}  - 2\,,
\end{align}
and $\si^\text{4FS}$ and $\si^\text{5FS}$ denote the total
inclusive cross section in the 4FS and the 5FS, respectively.
The theoretical uncertainties  in the 4FS and the 5FS calculations
should be added linearly, using the
weight $t$. In this way it is ensured that the combined
error is always larger than the minimum of the two individual
errors~\cite{Harlander:2011aa}:
\begin{align}
\De\si_\pm &= \frac{\De\si_\pm^\text{4FS}
  + t\,\De\si_\pm^\text{5FS}}{1+t}\,,
\end{align}
where $\De\si_\pm^\text{4FS}$ and
$\De\si_\pm^\text{5FS}$ denote the upper/lower uncertainty limits
of the 4FS and the 5FS, respectively.

\medskip
An up-to-date determination of the next-to-leading order total cross
section in the type~II 2HDM as a function of $\MHp$ and $\tb$ has
recently been presented in 
\citere{Flechl:2014wfa}, which constitutes the official recommendation
of the LHCHXSWG for heavy charged Higgs bosons. Also included in
\citere{Flechl:2014wfa} is an estimate of the theoretical uncertainties
due to missing higher-order corrections, parton distribution functions
and physical input parameters. 
Predictions in the 4FS and 5FS were compared and reconciled through
a recently proposed scale-setting prescription. 
Applying the Santander matching the ``best'' cross section prediction
for heavy charged Higgs bosons at the LHC is provided.

An interim recommendation of the LHCHXSWG on the evaluation of cross
sections and branching ratios in the 2HDM has been presented in
\citere{thdm-lhchxswg-reco}, however, with a focus on neutral Higgs
bosons. The two codes recommended for the Higgs boson decays, 
{\tt Hdecay}~\cite{hdecay} and {\tt 2HDMC}~\cite{2hdmc} also include the
evaluation of charged Higgs boson decays in types~I-IV.


\section{Charged Higgs bosons in the MSSM}
\label{sec:mssm}

While the MSSM contains (at the tree-level) a 2HDM type~II, due to
Supersymmetry (SUSY), special relations are enforced, and via loop
corrections the full SUSY spectrum enters the predictions.

The Higgs sector of the MSSM
contains two Higgs doublets, leading to five
physical Higgs bosons. At tree-level these are the light and heavy
$\cp$-even $h$~and $H$, the $\cp$-odd $A$ and the charged $H^\pm$. At
lowest order the Higgs sector can be described besides the SM
parameters by two additional independent
parameters, chosen to be the mass of the $A$~boson, $\MA$ 
(in the case of vanishing complex phases) and $\tb$. 
Accordingly, all other masses and couplings can be predicted at
tree-level, e.g.\ the charged Higgs boson mass
\begin{align}
\label{MHptree}
\mHp^2 &= \MA^2 + \MW^2~.
\end{align}
$M_{Z,W}$ denote the masses of the $Z$~and $W$~boson,
respectively. 
This tree-level relation receives higher-order corrections, where the loop
corrected charged Higgs-boson mass is denoted as $\MHp$. 
Three codes exist for the calculation of $\MHp$ and the various decay
widths, 
{\tt FeynHiggs}~\cite{feynhiggs,mhiggslong,mhiggsAEC,mhcMSSMlong,mhcMSSM2L,Mh-logresum}, 
{\tt CPsuperH}~\cite{cpsh}
and {\tt Hdecay}~\cite{hdecay}.

The relation between the bottom-quark mass and the Yukawa coupling
$h_b$, which controls also the interaction between the Higgs fields and
the sbottom quarks, is affected by higher-order corrections,
summarized in the quantity $\db$~\cite{deltamb1,deltamb2,db2l}.
These, often called threshold corrections, are generated either by
gluino--sbottom one-loop diagrams (resulting in \order{\alb\als}
corrections), or by chargino--stop loops (giving
\order{\alb\alt} corrections). 
The effective Lagrangian for the charged Higgs is given by~\cite{deltamb2}
\begin{align}
\label{effL}
\cL \sim V_{tb} 
      \KKL \KL \frac{\mbms}{1 + \db} \, \tb + \frac{\mt}{\tb} \KR
           H^+ \bar{t}_L b_R \KKR + {\rm h.c.}
\end{align}
Here $V_{tb}$ denotes the $(3,3)$ element of the CKM matrix, $\mbms$ is
the running bottom quark mass, and $\mt$ is the top quark mass.
Analytically one finds $\db \propto \mu \tb$, where $\mu$ is the Higgs
mixing parameter, which is (generally) of the same size as SUSY mass
scales. 
Large positive (negative) values of $\db$ lead to a strong suppression
(enhancement) of the bottom Yukawa coupling.

\begin{figure}[htb!]
\begin{center}
\includegraphics[width=0.45\textwidth]{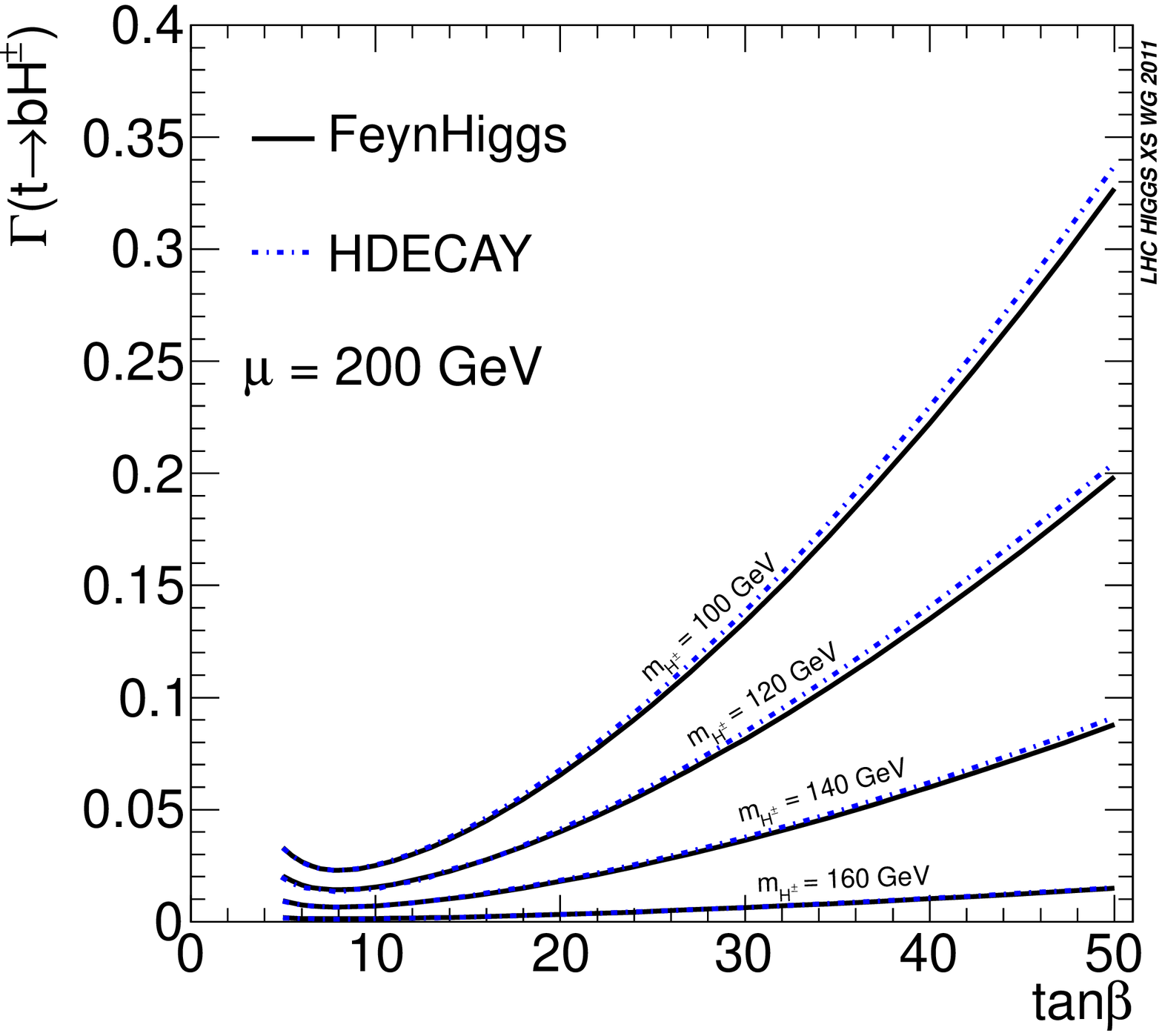}
\includegraphics[width=0.45\textwidth]{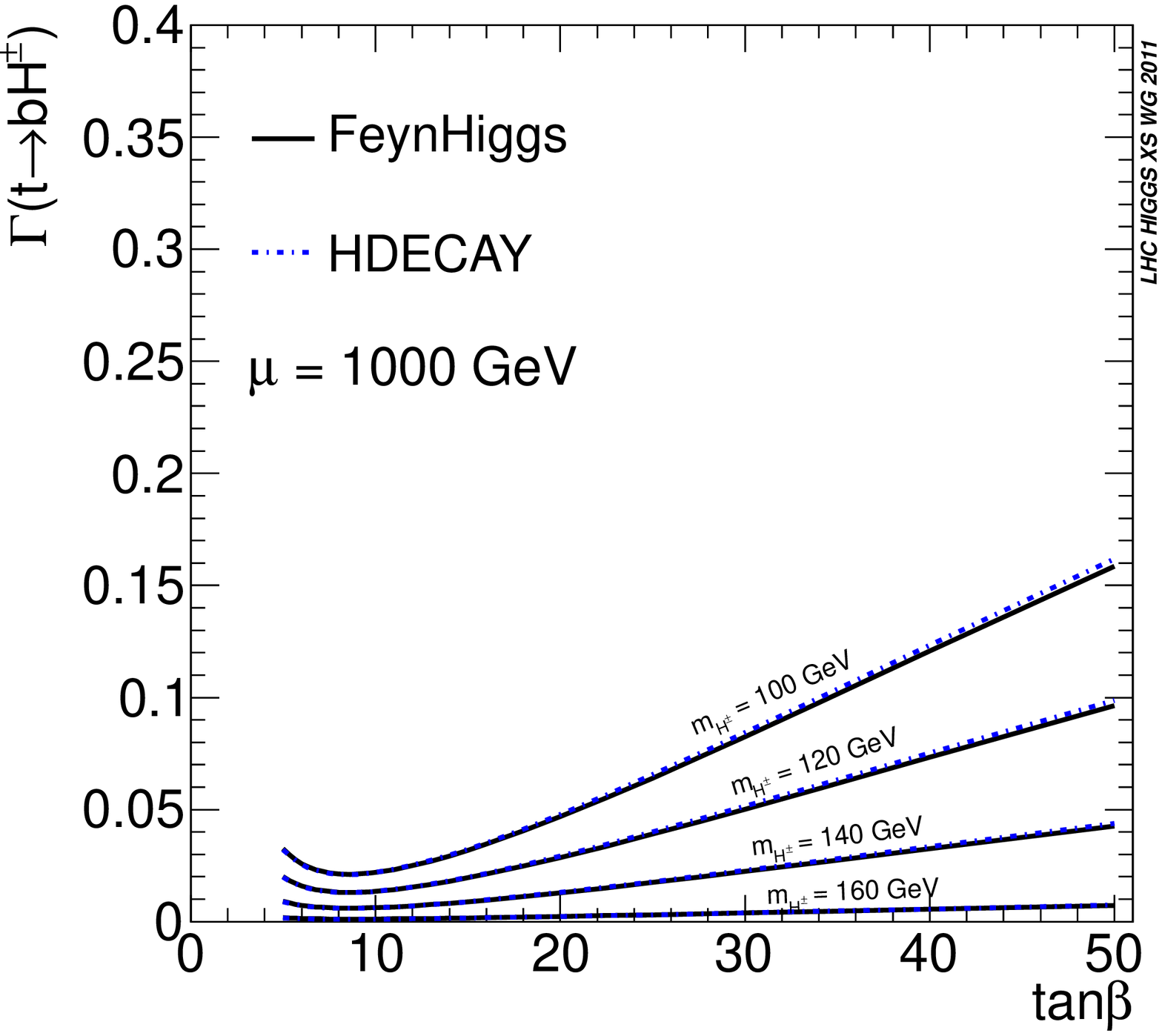}\\
\includegraphics[width=0.45\textwidth]{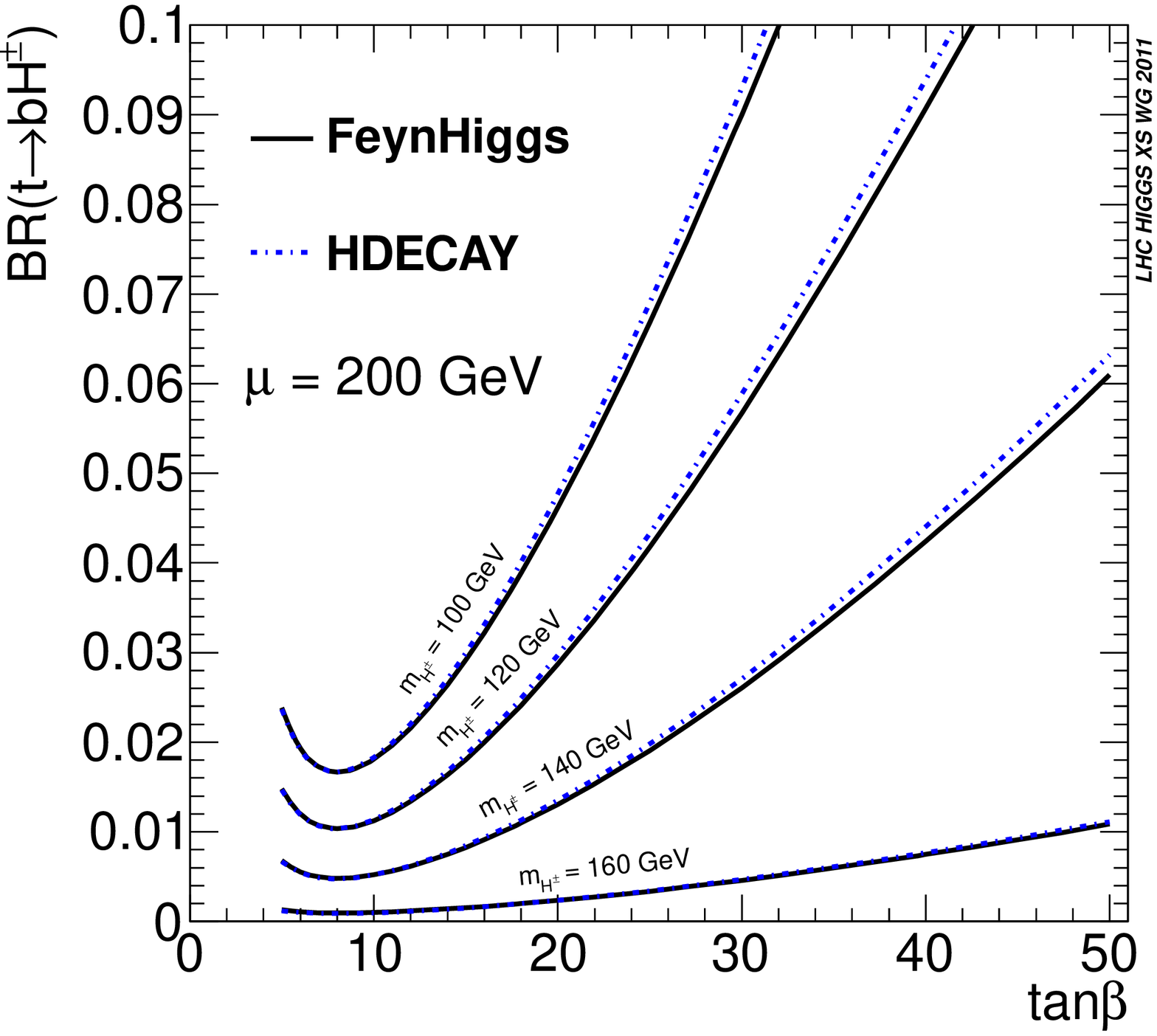}
\includegraphics[width=0.45\textwidth]{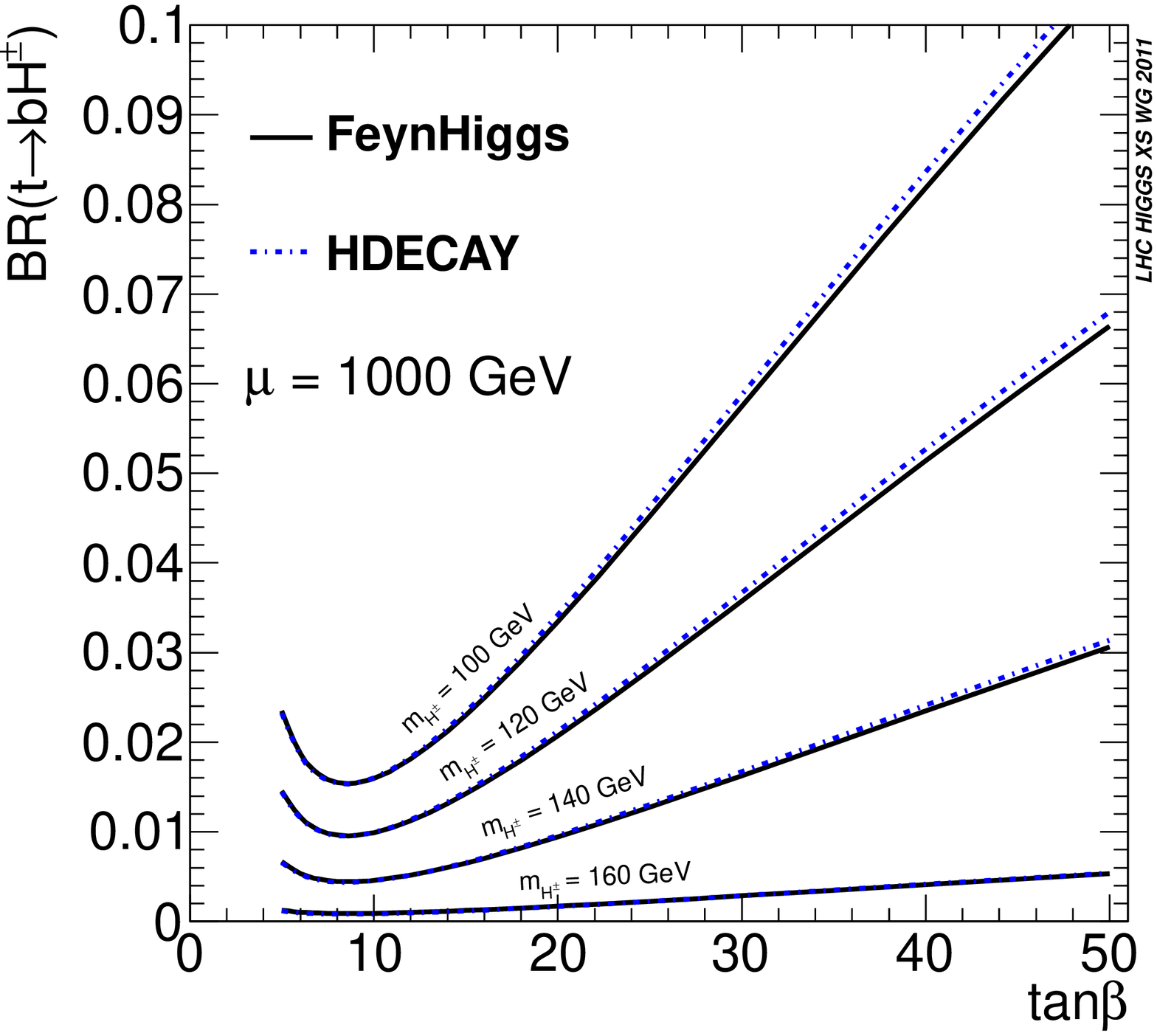}
\caption{Comparison of the $\Ga(t \to H^+ b)$ (upper row) and 
$\br(t \to H^+ b)$ (lower row) between {\tt FeynHiggs} and {\tt Hdecay}.
The results are shown for various values of $\MHp$ and for 
$\mu = 200 (1000) \gev$ in the left (right) column (taken from \citere{YR2}).
}
\label{fig:lightHiggsComp}
\end{center}
\end{figure}

\medskip
For the evaluation of the light charged Higgs production cross section
the decay $t \to H^\pm b$ has to be evaluated including SUSY loop
corrections, where the main contribution stems from \refeq{effL}. The
LHCHXSWG compared the codes {\tt FeynHiggs} and {\tt Hdecay} as shown in
\reffi{fig:lightHiggsComp}~\cite{YR2}.
The top row shows the decay width, while the bottom row contains the
result for the branching ratios. The parameters are chosen according to
the $\mhmax$~scenario~\cite{benchmark2} with $\mu$ set to $200 (1000) \gev$
in the left (right) column. One can see that the agreement between the
two codes, despite some differences in the $\db$ evaluation (see
\citere{YR2} for details) is excellent.

\begin{figure}[htb!]
\begin{center}
\includegraphics[width=0.70\textwidth]{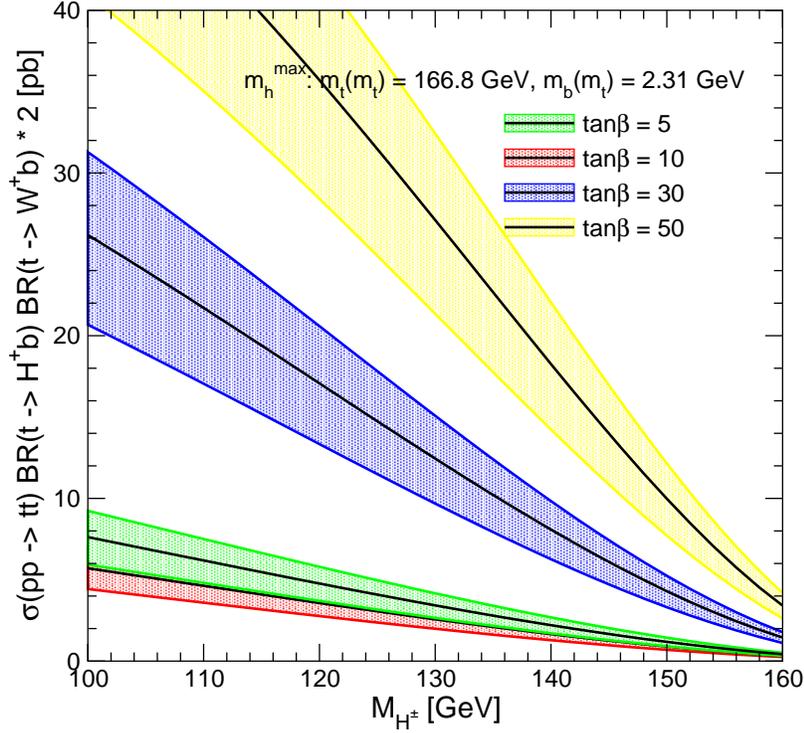}
\caption{$\si_{tt} \cdot \br(t \to b H^\pm) \cdot \br(t \to b W^\pm) \cdot 2$ 
including scale and PDF uncertainties, uncertainties for missing
electroweak and QCD corrections, and $\db$-induced uncertainties for
$\sqrt{s} = 7 \tev$ (taken from \citere{YR2}).
}
\label{fig:totalUnc}
\end{center}
\end{figure}

The LHCHXSWG also estimated 
the overall uncertainty of the light charged Higgs production, evaluated
in the $\mhmax$ scenario. The result is shown in
\reffi{fig:totalUnc}~\cite{YR2}, where
\begin{align}
\si_{tt} \cdot \br(t \to b H^\pm) \cdot \br(t \to b W^\pm) \cdot 2
\end{align}
is shown for $\sqrt{s} = 7 \tev$ 
as a function of $\MHp$. The uncertainty estimate combines the
accuracies for the top quark production (parametric and intrinsic
uncertainty) and for the top quark decay (including intrinsic
uncertainties on $\db$). 
The result is shown for $\tb = 5, 10, 30, 50$. 
As can be seen, the uncertainties are still substantial. They have to be
taken into account for reliable and robust bounds on the MSSM parameter
space from the non-observation of a light charged Higgs. Conversely,
using a potential observation of a light charged Higgs for a
determination of the underlying parameters would require a substantial
reduction of the uncertainties.

\medskip
The LHCHXSWG also provides branching ratio predictions for the MSSM
Higgs bosons, including the charged Higgs boson.
The procedure adopted by the LHCHXSWG goes as follows.
After the calculation of Higgs-boson masses and mixings from the
original SUSY input, a
combination of the results from {\tt Hdecay} and 
{\tt FeynHiggs} on the various decay channels is performed to 
obtain the most accurate result for the branching ratios currently
available. (For the general procedure, see \citere{BR}.) In a first
step, all partial widths have been calculated as accurately as
possible. Then the branching ratios have been derived from this full
set of partial widths. 
Concretely, {\tt FeynHiggs} was used for the evaluation of the
Higgs-boson masses and couplings from the original input
parameters, including corrections up to the two-loop level. 
The status of the various evaluations in {\tt FeynHiggs} and 
{\tt Hdecay} are detailed in \citere{YR2}.
The total decay width of the charged Higgs bosons is calculated as,
\begin{align}
\Gamma_{H^\pm} &= 
\phantom{+}  \Gamma^{\mathrm{FH}}_{H^\pm \to \tau \nu_\tau} 
+ \Gamma^{\mathrm{FH}}_{H^\pm \to \mu \nu_\mu} 
+ \Gamma^{\mathrm{FH}}_{H^\pm \to h W^\pm} 
+ \Gamma^{\mathrm{FH}}_{H^\pm \to H W^\pm} 
+ \Gamma^{\mathrm{FH}}_{H^\pm \to A W^\pm} \nonumber\\
&\quad 
+ \Gamma^{\mathrm{HD}}_{H^\pm \to tb}
+ \Gamma^{\mathrm{HD}}_{H^\pm \to ts}
+ \Gamma^{\mathrm{HD}}_{H^\pm \to td}
+ \Gamma^{\mathrm{HD}}_{H^\pm \to cb}
+ \Gamma^{\mathrm{HD}}_{H^\pm \to cs}
+ \Gamma^{\mathrm{HD}}_{H^\pm \to cd} \nonumber \\
&\quad
+ \Gamma^{\mathrm{HD}}_{H^\pm \to ub}
+ \Gamma^{\mathrm{HD}}_{H^\pm \to us}
+ \Gamma^{\mathrm{HD}}_{H^\pm \to ud}~,
\end{align}
followed by a corresponding evaluation of the respective branching
ratio. Decays to strange quarks or other lighter fermions have been neglected. 

\begin{figure}[htb!]
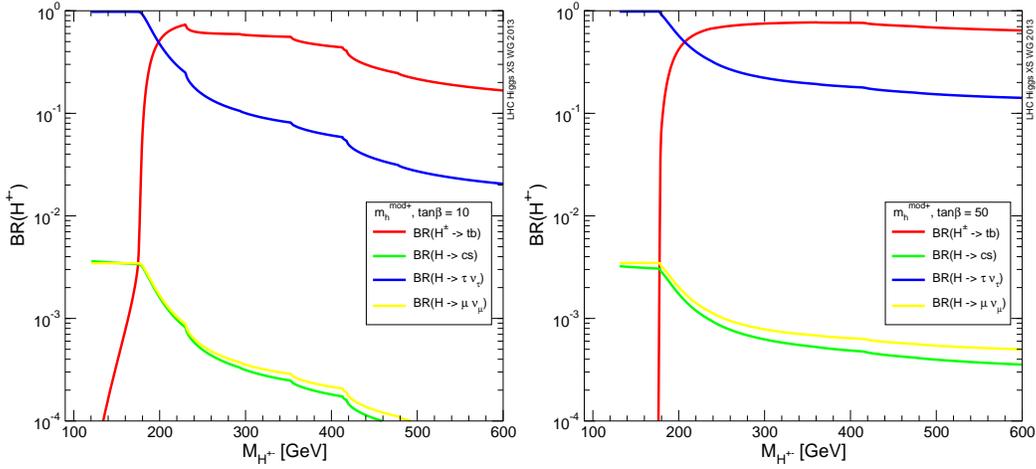

\begin{center}
\includegraphics[width=0.45\textwidth]{YRHXS3_BR_fig33}
\includegraphics[width=0.45\textwidth]{YRHXS3_BR_fig34}
\caption{Charged Higgs boson branching ratios in the 
$\mh^{\rm mod+}$ scenario~\cite{benchmark4} for $\tb = 10 (50)$ in the
  left (right) plot as a function of $\MHp$ (taken from \citere{YR3}).
}
\label{fig:BRHp}
\end{center}
\end{figure}

Example results in the $\mh^{\rm mod+}$ scenario~\cite{benchmark4} are
given in \reffi{fig:BRHp}~\cite{YR3}. The left (right) plot show the BRs
for $\tb = 10 (50)$ as a function of $\MHp$. The various kinks visible
in the left plot  stem from the decay channels to a
chargino/neutralino pair, which are not explicitely included into the BR
predictions yet.

\section{Conclusions}

The LHCHXSWG forms an important part of the efforts to identify the
mechanism of EWSB at the LHC. Among many other activities, it provides cross
sections and branching ratios for charged Higgs bosons as they are
predicted by the 2HDM and/or the MSSM. Here we briefly reviewed some of
the predictions for light and heavy charged Higgs bosons, including
evaluations of the respective uncertainties.


\subsection*{Acknowledgements}

We thank the organizers of C$H^{\mbox{}^\pm}$\hspace{-2.5mm}arged 2014
for the invitation and the, as always, pleasant and productive atmosphere.
We thank particularly many members of the LHCHXSWG, who obtained the
results reviewed here.
The work of S.H.\ is supported in part by CICYT 
(grant FPA 2013-40715-P) and by the Spanish MICINN's Consolider-Ingenio 
2010 Program under grant MultiDark CSD2009-00064.


\end{document}